\documentclass[11pt,a4]{article}
\usepackage{amsmath}
\usepackage{amssymb}
\usepackage{mathabx}
\usepackage{amsfonts}
\usepackage{array}
\usepackage{graphicx}
\usepackage{mathrsfs}
\usepackage{multirow}
\usepackage{siunitx}
\usepackage{cite}
\usepackage{amsmath}
\usepackage{algorithmic}
\usepackage{array}
\usepackage{graphicx}
\usepackage{epstopdf}
\usepackage{color,soul}
\newcommand{\bea}{\begin{eqnarray*}}
\newcommand{\eea}{\end{eqnarray*}}
\newcommand{\beao}{\begin{eqnarray}}
\newcommand{\eeao}{\end{eqnarray}}

\begin{document}
\title{Piezoelectric properties of substitutionally doped $\beta$-Ga$_2$O$_3$}

\author{Lijie Li}

\date{}

\maketitle
\thanks{College of Engineering, Swansea University, Swansea, Bay Campus, SA1 8EN, UK. e-mail: L.Li@swansea.ac.uk} 
\begin{abstract}
Modern semiconductor materials are increasingly used in multidisciplinary systems demonstrating cross-interactions between mechanical strains and electronic potentials, which gives rise to ubiquitous applications in high sensitivity, self-powered sensor devices. One of fundamental prerequisites for such semiconductor materials to exhibit piezoelectric properties is the noncentrosymmetry of the crystal structures. $\beta$-Ga$_2$O$_3$ has been an emerging compound semiconductor material due to its ultra-wide bandgap. However the pristine $\beta$-Ga$_2$O$_3$ has an inversion center, displaying no piezoelectric effect. This work discovered that substitutionally doped $\beta$-Ga$_2$O$_3$ possesses piezoelectric property by using first principles method, while majority of previous research on its substitutional doping has been focusing on the purposes of increasing electrical conductivity and formation of the semiconductor heterojunctions. More interestingly, it is unveiled from this work that the formation energy has a clear relation with the piezoelectric coefficient. 
\end{abstract}


\section{Introduction}

Gallium oxide (Ga$_2$O$_3$) has been increasingly investigated due to its ultra-wide bandgap (4.8 eV), which has sparked enormous research and development activities especially in areas of power electronics and solar blind UV photodetection \cite{Zhang_APL,Li_GaO}. Ga$_2$O$_3$ has five crystallographic phases, $\alpha$, $\beta$, $\gamma$, $\delta$, and $\epsilon$. Among these phases, $\beta$ type is the most thermodynamically stable. It has a monoclinic crystal structure with Ga in the tetrahedral and octahedral sites \cite{Liandi}. On the contrary, wide bandgap gives large resistivity, which hinders the material from being implemented in electronic devices. Moreover, often special semiconductor types (i.e. p- and n- type) are required in the heterojunctions, which demands that $\beta$-Ga$_2$O$_3$ has to be intentionally doped. Here in this work, it is focused on the substitutional doping. Many research activities surrounding $\beta$ type have been doping of the pristine $\beta$-Ga$_2$O$_3$ with various metals attempting to increase its conductivity, so that it can be used to form diodes/transistors in practical applications \cite{SPIE}.  Both n-type and p-type $\beta$-Ga$_2$O$_3$ have been achieved through substitutionally doping various metals such as using Sn doping to realize n-type \cite{CEC}, and using Zn doping to achieve p-type \cite{WANG}. Strain induced polarization in modern semiconductor materials has attracted much attention due to its promising application in tuneable devices with improved performances \cite{Wu_nature, Zhangyan}. In order for the semiconductor materials to be piezoelectrically tuneable, their crystal structures must display noncentrosymmetric property. Although $\beta$-Ga$_2$O$_3$ is superior in many applications, it doesn’t  exhibit piezoelectricity. So far there is very few reports on piezoelectricity of Ga$_2$O$_3$, there is one paper published recently reporting piezoelectric property of $\epsilon$-Ga$_2$O$_3$ \cite{Guo} based on first principles analysis, which is the only crystallographic phase that does not have an inversion symmetry among all Ga$_2$O$_3$ phases. First principles method has been widely used to study new material effects previously \cite{Ma_Jia, Mengle}. Here for the first time, it is unveiled that the substitutionally doped $\beta$-Ga$_2$O$_3$ possess piezoelectric properties due to broken inversion symmetry caused by substitutionally doping with metal atoms. Detailed first principles analysis and discussion are described to elucidate this hypothesis.

\begin{figure}[!t]
\centering
\includegraphics[width=3in]{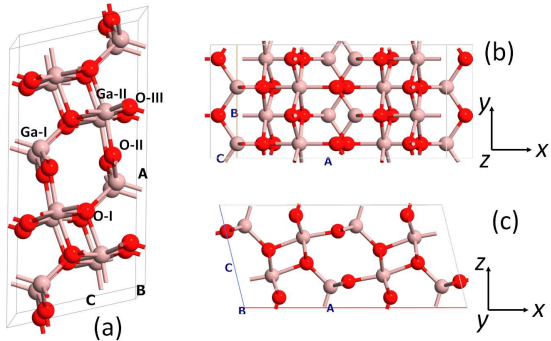}
\caption{Crystal model of the $\beta$-Ga$_2$O$_3$ supercell containing 16 Ga atoms and 24 O atoms. (a) 3D atomic structure showing different types of Ga and O locations. (b) atomic structure in $x$-$y$ plane and $x$-$z$ plane (c).}
\label{fig1}
\end{figure}

\section{Analysis and Results}
In the simulations, monoclinic supercell of $\beta$-Ga$_2$O$_3$ has been built with 16 Ga atoms and 24 O atoms according to the crystal parameters in Ref.\cite{Geller} (Figure 1). The QuantumATK software has been used for the simulation \cite{ATK}. The supercell was then optimized using the generalized gradient approximations (GGA) and Perdew–Burke–Ernzerhof (PBE) exchange-correlation functional. The density mesh cut-off was set to 125 Hartree and k-point sampling was 4$\times$4$\times$4. In the geometry optimization setting, the force tolerance was 0.03 eV/ \r{A}, and stress tolerance was set to 0.001 eV/ \r{A}$^3$. The optimized cell dimension is detailed in Table 1. The space group of the optimized $\beta$-Ga$_2$O$_3$ supercell is C2/m (group 12), which has an inversion centre. The unit cell of the $\beta$-Ga$_2$O$_3$ consists of two crystallographically different Ga atoms, one with tetrahedral (Ga-I) and the other with octahedral (Ga-II) coordination geometry. In addition it contains three types of oxygen atoms (O-I, O-II, and O-III) \cite{Zhang_APL}. Before investigating the doped $\beta$-Ga$_2$O$_3$, analysis on the pure $\beta$-Ga$_2$O$_3$ has been conducted to validate the simulation method by comparing simulated bandstructure and density of state with previous reported. It has been well known that the GGA-PBE functional usually leads to underestimation of the bandgaps, therefore a more accurate algorithm, Meta-GGA (MGGA) that takes account of local density $\rho (r)$ and the gradient of the density $\nabla \rho (r)$, as well as the kinetic-energy density $\tau (r)$. The difference between GGA and MGGA is that the latter considers $\tau (r)$. Mathematically the exchange potential $v_x^{TB} (r)$ is expressed as \cite{ATK}
\begin{equation}
    v_x^{TB} (r)=cv_x^{BR} (r)+\frac{3c-2}{\pi}\sqrt{\frac{4\tau (r)}{6\rho (r)}} 
\end{equation}
where $\tau (r) = 1/2 \sum_{i=1}^N |\nabla \psi _i (r)|^2$,$\psi _i (r)$ is the i’th Kohn-Sham orbital, and $v_x^{BR} (r)$ is the Becke-Roussel exchange potential. The parameter $c$ can be extracted by fitting to the experimentally obtained bandgap. Here in this work, $c$ parameter was selected as 1.4 based on which the calculated bandgap for the $\beta$-Ga$_2$O$_3$ is 4.8 eV. This bandgap is consistent with the value reported in experiments. MGGA method has been used in calculating bandstructures and optical properties. MGGA is one of generally adopted methods to correct the underestimated bandgap by GGA \cite{Tran}, which gives reasonable precisions as some computationally expensive methods such as hybrid functionals (HSE). Except for the bandgap value, GGA is an appropriate method for other properties. In this work, MGGA was used to calculate the bandgap and optical properties as the optical properties are mainly dependent on the bandstructures. GGA was used to calculate the formation energies and piezoelectric coefficients, which has been considered as an appropriate approach \cite{APL_2002} \cite{Tang}.

\begin{figure*}[!t]
\centering
\includegraphics[width=6in]{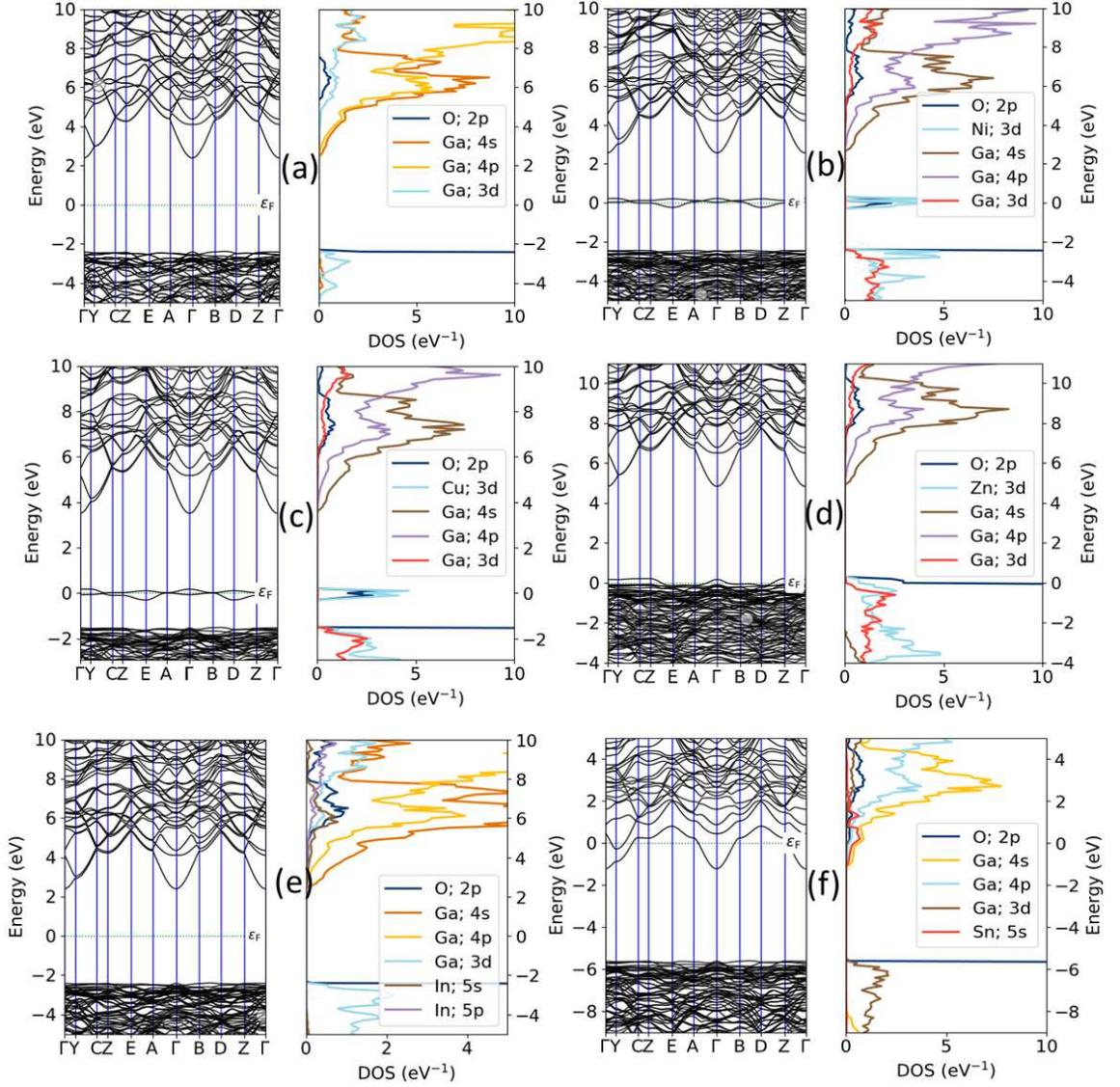}
\caption{Simulated band structures and electron density of states for various substitutionally doped $\beta$-Ga$_2$O$_3$ crystal structures: (a) intrinsic $\beta$-Ga$_2$O$_3$; (b) doped by Ni; (c) doped by Cu; (d) Zn doped; (e) In doped; and (f) Sn doped.}
\label{Fig_2}
\end{figure*}

The electrical conductivity of intrinsic $\beta$-Ga$_2$O$_3$ is very low. As seen from Figure 2a, the valence band maximum is mainly formed by fully filled O 2p states, and the conduction bands in proximity to the fermi level are mainly occupied by the Ga 4s and 4p orbitals. To increase the electrical conductivity, $\beta$-Ga$_2$O$_3$ is usually doped by various metals, inducing additional donor or acceptor levels, corresponding to n- or p- type respectively. Most notably, Zinc (Zn) doping leads to increased acceptor levels (p-type conductivity) \cite{WANG}, and Tin (Sn) doping causes increased n-type conductivity. It was reported that the resistivity of the Sn doped $\beta$-Ga$_2$O$_3$ film decreased from 1.79 $\times$ 10$^7$ Ohm-cm to 0.12 Ohm-cm as the Sn concentration increases from 0 to $10\%$ \cite{Du}. Here in this work, it is focused on the metals that are located close to the Gallium on the Periodic table, i.e. Ni (28), Cu (29), Zn(30) on the same row with Ga (31), In (49) and Sn (50) on the next row, as those elements have similar numbers of electrons in the outer orbitals. Only the substitutional doping is considered in this work, which means the doping is implemented by replacing Ga atoms with one of metals mentioned above. It was well reported that most metal atoms take the Ga-II position as substitution position at Ga-II has a lower formation energy than that of Ga-I \cite{Tang}. However it is opposite for Zn atom \cite{ZHANG_PB}. Hence in the models, all types of metals are taking position of Ga-II except for Zn. Precisely the doped supercell is composed of 24 O atoms, 15 Ga atoms and one other metal atom, corresponding to 6.25 at$\%$ concentration. This doping concentration has been used in theory, which was compared with experiments \cite{YZhang}. In \cite{Imura}, 8 at$\%$ Sn doping into Ga$_2$O$_3$ was used experimentally. MGGA simulation has been conducted for the substitutional doped $\beta$-Ga$_2$O$_3$ for band structures and projected density of states. From Figures 2b and 2c, Ni and Cu doping made very similar impact on the bandgap, i.e. there are induced donor bands contributed by Ni 3d and Cu 3d orbitals and the bandgaps remain similar to the undoped case, around 5 eV. This is understandable as Ni and Cu are adjacent to each other on the periodic table and having similar outer orbitals. With regards to the Zn doping, Zn atom takes the tetrahedral Ga position, inducing a shallow accepter energy level contributed by Zn 3d (Figure 2d), which clearly displays a p-type conductivity. This is in consistent with previous experimental observation \cite{Li_cgd}. In doping makes very little impact on bandstructures, i.e. bandgap has been narrowed by around 0.1 eV (Figure 2e). Sn doped $\beta$-Ga$_2$O$_3$ has been very popular and it induces n-type conductivity, which is attributed to the Sn 5s orbitals shown in Figure 2f. 

Optical properties of the pure $\beta$-Ga$_2$O$_3$ and doped  $\beta$-Ga$_2$O$_3$ such as absorption coefficient and refractive index can be derived from the DFT calculations. The detailed procedure of these derivation process was described in \cite{Li_JCE, Li_ASS}. Briefly the calculation started with the susceptibility tensor that determines the polarisation level of the material due to an electric field. The complex dielectric constant was then deduced from the susceptibility tensor. The refractive index is related to the complex dielectric constant, and the optical absorption is related to the imaginary part of the dielectric constant.

\begin{figure}
\centering
\includegraphics[width=3.6in]{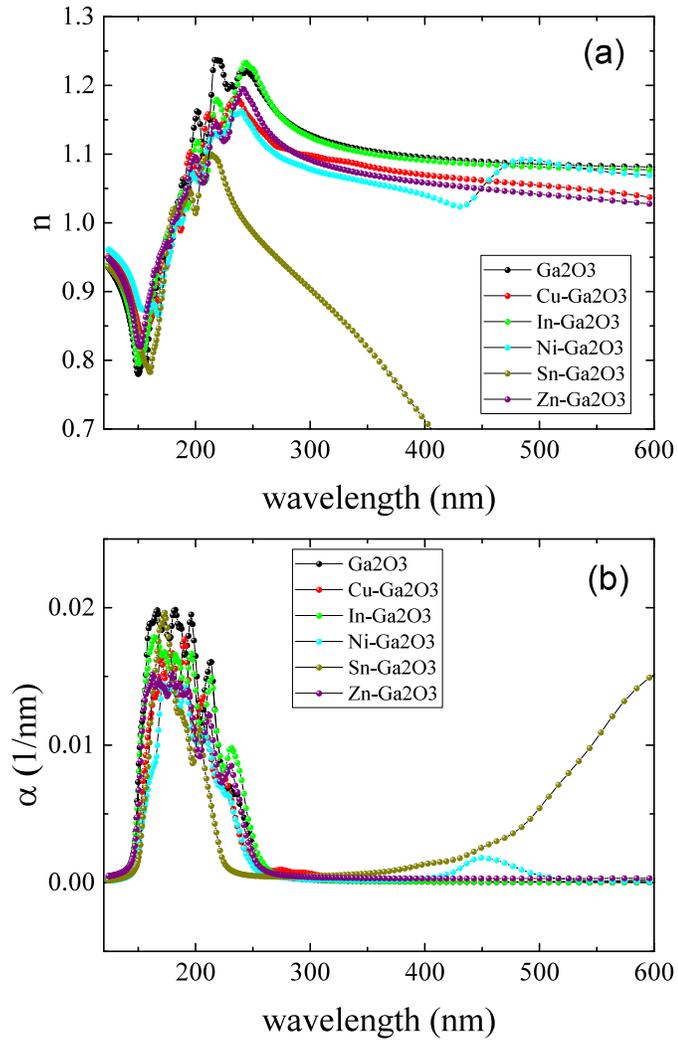}
\caption{Calculated optical properties of doped $\beta$-Ga$_2$O$_3$. (a) calculated refractive index versus wavelength; (b) calculated absorption coefficient versus wavelength.}
\end{figure}

The calculated refractive indices against wavelength for all doping configurations are very similar except for the Sn doped that demonstrates significant differences from others. With regards to the absorption coefficient, the various doped $\beta$-Ga$_2$O$_3$ in the Figure 3b show that the peak magnitudes of all cases under investigation at around 200 nm wavelength show an decreasing trend from the pure to doped $\beta$-Ga$_2$O$_3$, especially for the Zn and In doped cases. Nevertheless, all peak values in the band of 150nm to 250 nm are between 0.015 – 0.02 nm$^{-1}$. That is matching with the optical bandgaps for all these crystal structures. Interestingly it shows a graduate increase for Sn doped $\beta$-Ga$_2$O$_3$ above 350 nm up to 600 nm wavelength.

\begin{table*}
\centering
\caption{Optimized lattice constants for $\beta$-Ga$_2$O$_3$ and its doped crystals.}
\begin{tabular}{|c|c|c|c|c|c|c|}
\hline
 &$\beta$-Ga$_2$O$_3$&Ni-Ga$_2$O$_3$&Cu-Ga$_2$O$_3$&
Zn-Ga$_2$O$_3$&In-Ga$_2$O$_3$&Sn-Ga$_2$O$_3$\\
\hline
$a$& 12.427 \AA & 12.384 \AA & 12.415 \AA &12.496 \AA
& 12.514 \AA & 12.553 \AA \\
\hline
$b$& 6.179 \AA & 6.171 \AA & 6.177 \AA &6.189 \AA
& 6.233 \AA & 6.247 \AA \\
\hline
$c$& 5.885 \AA & 5.881 \AA & 5.886 \AA &5.868 \AA
& 5.910 \AA & 5.942 \AA\\
\hline
$\beta$& 103.648$^{\circ}$ & 103.589$^{\circ}$ & 103.580$^{\circ}$ &103.472$^{\circ}$
& 103.481$^{\circ}$ & 103.614$^{\circ}$ \\
\hline
\end{tabular}
\end{table*}

\begin{table*}
\centering
\caption{Lattice type and space group for $\beta$-Ga$_2$O$_3$ and its doped crystals}
\begin{tabular}{|c|c|c|c|c|c|c|}
\hline
&$\beta$-Ga$_2$O$_3$&Ni-Ga$_2$O$_3$&Cu-Ga$_2$O$_3$&
Zn-Ga$_2$O$_3$&In-Ga$_2$O$_3$&Sn-Ga$_2$O$_3$\\
\hline
Bravais Lattice& Monoclinic & Monoclinic & Monoclinic & Monoclinic
& Monoclinic & Monoclinic \\
\hline
Space group& C2/m & Pm & Pm & Pm
& Pm & Pm \\
\hline
Piezoelectric& No & Yes & Yes & Yes
& Yes & Yes\\
\hline
\end{tabular}
\end{table*}

Table I shows calculated optimized crystal dimensions for various doping conditions. The geometric optimization of substitutionally doped crystals followed the procedure described in Ref.\cite{Li_ASS}. It is noted that the GGA in conjunction with PBE has been used in geometrical optimization procedure to relax the crystal structures in order to achieve minimum stress inside the crystal. Shown from the Table I, slight variations appear for different doping configurations. For example, Cu and Ni doped $\beta$-Ga$_2$O$_3$ have slightly reduced dimensions, while Zn, In, and Sn doped $\beta$-Ga$_2$O$_3$ have increased crystal lattice parameters. Interestingly from the periodic table, Ni, Cu and Zn have smaller atomic numbers than Ga. In and Sn have larger atomic numbers than Ga. Moreover calculation has been conducted to arrive at the space group of the pristine $\beta$-Ga$_2$O$_3$ and doped cases. All cases investigated here are monoclinic crystals (Table II). It is found that the space group of the pristine $\beta$-Ga$_2$O$_3$ is C2/m, which is centrosymmetric, displaying no strain induced polarisation. This has been well reported in literatures \cite{Geller}. However, looking at the space groups of doped $\beta$-Ga$_2$O$_3$, they are all Pm type, which does not have a center of symmetry. Clearly this new finding leads to a plausible conclusion that the substitutionally doped $\beta$-Ga$_2$O$_3$ will have piezoelectric properties, i.e. charges can be generated from mechanical strains. Further calculation has been conducted subsequently to obtain piezoelectric coefficients of doped $\beta$-Ga$_2$O$_3$.

\begin{figure}
\centering
\includegraphics[width=4in]{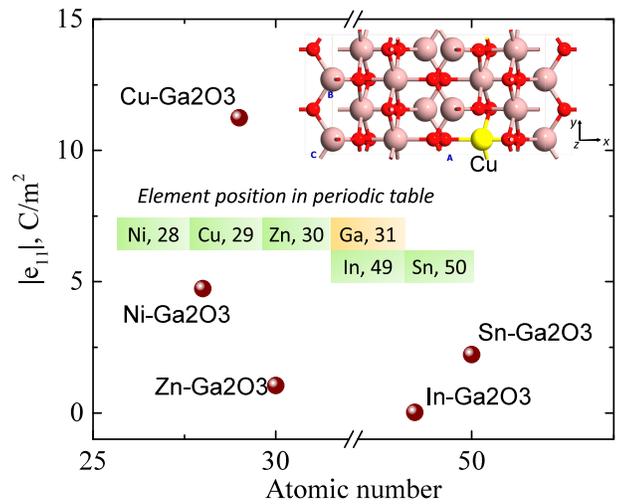}
\caption{Piezoelectric coefficient for doped $\beta$-Ga$_2$O$_3$. Insert graph shows doping elements in periodic table. Simple Cu doped supercell is also shown on the top right corner of the figure. }
\end{figure}

\begin{figure}
\centering
\includegraphics[width=4in]{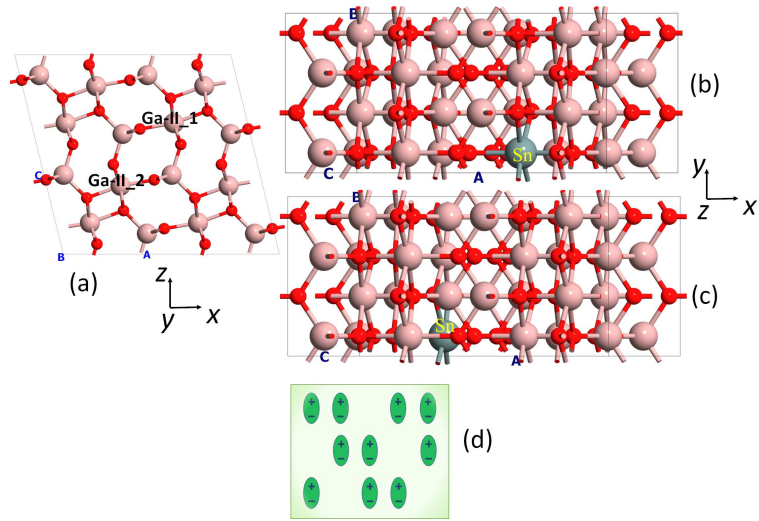}
\caption{Bigger model containing 80 atoms. (a) model showing two possible Ga-II locations. (b) Sn doping in the location 1, and (c) location 2. (d) Schematic illustration to show that the material will still demonstrate piezoelectricity considering imperfectness and randomness of the doping in experiments.}
\end{figure}

Identical crystals structures are used as in previous sections for piezoelectric coefficient calculation for pure  $\beta$-Ga$_2$O$_3$ and doped cases in order to achieve consistent simulation results. It is focused on the piezoelectric coefficient at '11' direction, as it is only the $x$ axis aligns with the direction $A$. The substitutionally Cu doped supercell in $x$-$y$ plane is shown in Figure 4 (insert). The strain induced polarization is calculated using a Berry-phase approach in the software and $e_{11}$ is calculated, which reflects the piezoelectric polarization along x-axis (Figure 4) subjecting to strain in x-axis. The calculated $e_{11}$ for all crystal cases are summarised in Figure 4. It is seen that Ni doped $\beta$-Ga$_2$O$_3$ has $e_{11}$ of 4.74 $C/m^2$, $e_{11}$ of Cu doped is 11.24 $C/m^2$. Zn doped has $e_{11}$ of 1.05 $C/m^2$. In and Sn doped have the values of 0.02 $C/m^2$ and 2.23 $C/m^2$ respectively.  After showing the element position of the doping metals in the periodic table (insert picture in Figure 4), one cannot help link the piezoelectric coefficients of all cases with their atomic numbers. Ni and Cu are farther away to the left of the Ga atom, while Sn is farthest to the right of Ga atom. These three doped crystals exhibit large piezoelectric coefficients. When Ga atom is substituted by closer metal atoms in the periodic table such as Zn and In, the piezoelectric coefficient is much smaller. 

\begin{figure}
\centering
\includegraphics[width=4in]{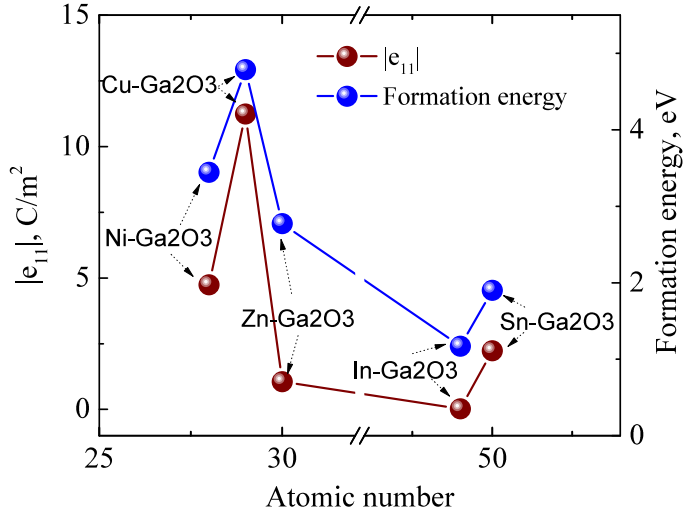}
\caption{Calculated formation energies in comparison with piezoelectric coefficients for various doping configurations.}
\end{figure}

Experimentally, detecting the exact location of diluted impurities in the $\beta$-Ga$_2$O$_3$ crystal lattice is challenging. One possibility is to measure the small changes in valence band density of states through X-ray photoelectron spectroscopy. Hence one may comment that that the first principles simulation on piezoelectric effect could be a speculation due to imperfectness or randomness of the substitutional doping in the practical scenarios. More DFT simulations have been conducted with a larger size supercell using Sn doped as example containing 80 atoms ($Ga_{31}SnO_{48}$), which is twice diluted compared with the previous simulations. Also the Sn has been placed randomly in two possible positions in Figure 5a (Ga-II\_1 and Ga-II\_2) in this supercell. The two possible dopings are shown in Figures 5b and 5c. The values of $e_{11}$ for these two doping cases are calculated to be 0.92 $C/m^2$ and 0.72 $C/m^2$ respectively. These values are approximately half of the value from the 40-atom model in Figure 4. Taking into consideration of imperfectness and randomness of doping in experiments, the calculations results show that the substitutional doped $\beta$-Ga$_2$O$_3$ will show piezoelectricity because the $e_{11}$ has the identical direction and will not cancel each other (schematically depicted in Figure 5d), but the actual experimental piezoelectric coefficient might be much smaller than calculated values due to reduced impurity levels.

The formation energy ($E_f$) for a substitutionally doped $\beta$-Ga$_2$O$_3$ by various metals is defined as 

\begin{equation}
E_f=E_{doped}-E_{bulk}+\mu_{Ga}-\mu_M
\end{equation}

where $E_{doped}$ is the total energy of the metal doped $\beta$-Ga$_2$O$_3$, $E_{bulk}$ is the total energy of the bulk material, $\mu_{Ga}$ and $\mu_M$ are the chemical potentials of Ga and dopant metal, respectively. Here only the Ga-rich condition is considered and GGA was used to calculate the formation energies. Although GGA-predicted values are slightly smaller than those by using HSE method, it will have little effect on comparison studies on various dopants\cite{Tang}. Comparisons of calculated formation energies and piezoelectric coefficients for different doping conditions are shown in the Figure 6. It is seen from Figure 6 that the formation energy is directly proportional to the piezoelectric coefficient. For instance, Cu doped has the largest formation energy as well as the largest piezoelectric coefficient. There is one exception where $e_{11}$ of Zn doped is slightly lower than Sn doped. That is possibly due to that Zn atom takes Ga-I position, while all the rest atoms take Ga-II position. From optimised supercells for different doping configurations, the cell volume (CV) looks like another determining factor. The CVs of Ni, Cu, Zn, In and Sn doped are 436.9, 438.7, 441.3, 448.2 and 452.8 $\AA^3$, while the bulk Ga$_2$O$_3$ has the CV of 439.2 $\AA^3$. On the same row with the Ga atom in the periodic table, the doping configurations with CVs closer to that of bulk Ga$_2$O$_3$ have larger piezoelectric coefficients. However, it is the opposite on the next row where In and Sn reside.

\section{Conclusion}
To summarize, density functional theory simulation has been conducted on doped $\beta$-Ga$_2$O$_3$. As opposed to undoped $\beta$-Ga$_2$O$_3$, substitutionally doped crystals do not have an inversion center, hence exhibiting piezoelectric performance. More study unveils that the formation energy of the doped configurations is relating to the magnitude of the piezoelectric coefficient. Larger formation energies of doped $\beta$-Ga$_2$O$_3$ will lead to stronger piezoelectricities. This investigation will hopefully serve as the tipping point in terms of applying doped $\beta$-Ga$_2$O$_3$ in the area of piezotronics.

\section*{Data Availability Statements}

The data that support the findings of this study are available from the corresponding author upon reasonable request.

\section*{Acknowledgment}

The author would like to thank...

\end{document}